# Transport Studies of Isolated Molecular Wires in Self-Assembled Monolayer Devices

V. Burtman, A. S. Ndobe, and Z. V. Vardeny*

Department of Physics, University of Utah, Salt Lake City, Utah 84112

*Abstract*

We have fabricated a variety of novel molecular diodes based on self-assembled-monolayers (SAM) of solid-state mixture (SSM) of molecular wires (1,4 benzene-dimethane-thiol; Me-BDT), and molecular insulator spacers (1-pentanethiol; PT) with different concentration ratios r of wires/spacers, which were sandwiched between two gold (Au) electrodes. We introduce two new methods borrowed from Surface Science to (i) confirm the connectivity between the Me-BDT molecules with the upper Au electrode, and (ii) count the number of isolated molecular wires in the devices.

The electrical transport properties of the SSM SAM diodes were studied at different temperatures via the conductance and differential conductance spectra. We found that a potential barrier caused by the spatial connectivity gap between the PT molecules and the upper Au electrode dominates the transport properties of the pure PT SAM diode (r = 0). The transport properties of SSM diodes with r-values in the range $10^{-8} < r < 10^{-4}$ are dominated by the conductance of the isolated Me-BDT molecules in the device. We found that the temperature dependence of the SSM diodes is much weaker than that of the pure PT device indicating the importance of the Me-BDT simultaneous bonding to the two Au electrodes that facilitate electrical transport. From the differential conductance spectra we also found that the energy difference, $\Delta$ between the Au electrode Fermi-level and the Me-BDT HOMO (or LUMO) level is ~1.5 eV; whereas it is ~2.5 eV for the PT molecule. The weak temperature dependent transport that we obtained for the SSM diodes reflects the weak temperature dependence of $\Delta$.

In addition, our measurements reveal that the conductance of SSM diodes scales linearly with r, showing that charge transport in these devices is dominated by the sum of the isolated Me-BDT molecular conductance in the device. Based on this finding, and the measured number of the Me-BDT molecules in the device we obtained the 'single molecule resistance', $R_M$. We measured $R_M = 6 \times 10^8$ $\Omega$ for isolated Me-BDT molecules, which is consistent with previous measurements using other transport measuring techniques. A simple model for calculating $R_M$, where the transport is governed by electron tunneling through the Me-BDT molecule using the WKB approximation, is in good agreement with the experimental data thus validating the procedures used for our measurements.

* To whom correspondence should be addressed; e-mail: val@physics.utah.edu

## I. INTRODUCTION

The emergence of new approaches for molecular engineering is a vital requirement for progress in molecular electronics, which should eventually lead to molecular devices with well-controlled properties and integrated circuits that compete with silicon technologies. In the last few years, a number of different experimental strategies have been used to probe electron transport through single conductive molecules (or molecular wires), such as electrode-molecule-electrode hetero-junctions, using, e.g., mechanical break junctions,[1] electro-migrated break junctions,[2] scanning nanoprobe microscopes[3,4] and crossed nanowires.[5] The initial insight gain on intramolecular transport processes in the new field of Molecular Electronics has subsequently led to many open questions. The pivotal difficulties in this young field are poor reproducibility of electrical transport of similar 'single molecule' devices from different research groups, which has led to controversies in the field;[6,7] and large discrepancies between theory and experiment.[8,9] These difficulties may be largely attributed to two issues: (i) absence of covalent bonding between the isolated molecules and one of the device metallic electrodes; and (ii) difficulty in determining the exact number of molecular wires in the 'single molecule' devices.[8,10] More specifically, there is a lack of experimental tools for verifying the electrical connectivity of single molecules to the electrodes. For example the available spectroscopic tools have limited capability of detecting the molecular-wire density or electrode/molecule bonding in the device, because of their miniature size. There is also lack of systematic transport studies for measuring single molecular resistance.[2, 11-20]

In this work we remedy this situation by suggesting a number of experimental procedures for checking the electrical connectivity and bonding of isolated conductive molecules to the electrodes, as well as counting the number of wire molecules in the device. For our studies we fabricated a variety of molecular diodes (or junctions) based on self-assembled-monolayers (SAM) of molecular wires (1,4-methane benzene dithiol; or Me-BDT) and molecular insulators (pentane 1-thiol; or PT) mixed together in different ratio concentrations, r. The SAM's were sandwiched between two gold electrodes, which are known for their inert properties. The resulting devices were characterized by a number of different spectroscopies for verifying the SAM growth and bonding to the gold electrodes. For r-values in the range of $r < 10^{-4}$ we found that the molecular wires are isolated in the otherwise insulator matrix, and thus the device electrical conduction is dominated by the 'single', isolated molecular wires down to $r = 5 \times 10^{-8}$. Below this r-value the device conductivity is limited by the finite conduction of the insulated matrix. We studied the I-V characteristics and differential conductance of devices having different r-values within the range $10^{-7} < r < 10^{-4}$, and verified the linear dependence of the device conductance with r in the regime dominated by the molecular wires. We also studied the temperature (T) dependent conductivity of the molecular diodes at various r-values, and obtained the changes in the electronic density of states of the molecular wire with T. In addition, we also introduced a surface titration approach to evaluate the surface density of molecular wires in the fabricated devices, and used this information to estimate the electrical resistance of a 'single' molecular wire isolated in a matrix of insulating molecules. We found that the electrical resistance, $R_M$ of isolated Me-BDT molecule in the PT matrix is $R_M \approx 6 \times 10^9 \, \Omega$ at small biasing voltage; this is in good agreement with a simple transport model based on the Landauer formula using the WKB approximation for tunneling through the molecule. The agreement between the data and the model calculation validates our method for counting the number of molecular wires in the device, as well as the assumption of charge transport dominated by isolated Me-BDT conducting molecules in the device.

## II. EXPERIMENTAL METHODS

**A. SAM growth**

The transport mechanism in low-dimensional organic structures is intimately related to the dimensionality of the underlying electronic system, which may transform with the molecular packing.[21-26] The question of how to manipulate the electronic dimensionality in organic structures, and the methods to detect it are not trivial. Here we used a molecular engineering approach in which SAM on metallic electrodes grows from solution mixtures of molecular wires (Me-BDT) and molecular insulating spacer molecules (PT) with a concentration ratio, $r = N_{BDT}/N_{PT}$, where $N_{BDT}$ and $N_{PT}$ are their respective molar concentrations. Our goal was to fabricate solid-state mixtures (SSM) with predictable structural features, where the isolated molecular wires are dispersed in the insulated matrix of aliphatic molecules that have 'dielectric' properties.

The devices were fabricated using the protocol shown schematically in Fig.1. The bottom Au electrode (about 30 nm thick) was deposited on a $SiO_2$/Si wafer using a DV-SJ/20C Denton Vacuum e-gun. The Me-BDT and PT precursors mixture was diluted with distilled toluene to 3 mM solution and air-free transferred to the home-built high-vacuum-based Shlenk line. The self-assembling process continued for about 12 hours in an argon atmosphere at room temperature. After the SAM growth was completed the samples were thoroughly washed in dry toluene and annealed in vacuum for 1 hour at 90ºC to remove any physisorbed precursors. The upper Au electrode was then evaporated through a shadow mask in a vertical cross electrodes configuration (Fig. 1C and 1D) using the DV-SJ/20C e-gun at 95ºC on the sample-holder. During self-assembly we varied the stoichiometric ratio, r, in the range $10^{-7} < r < 10^{-4}$. In this r-value range single molecular wires are isolated within the insulating PT matrix (see below) (Fig. 1B). Every Si chip contained three different devices; each with an active area of about 0.5 mm$^2$. The device concept is depicted in Fig. 1D.

Due to different SAM reaction rates, the ratio of the wire/insulator molecule density in the SAM configuration may not be equivalent to the stoichiometric ratio r in the solution. We assume that the wire and insulator molecules form solid-state mixture (SSM) in the monolayer, which is characterized by the nominal r-value from the solution mixtures. The actual density of molecular wires was determined by a surface titration method as described below. Changing the r-value in the solution thus tunes the conduction process in the SAM devices within the regime of charge transport through isolated Me-BDT molecules, namely $5 \times 10^{-8} < r < 10^{-4}$. The transport studies through 2D molecular aggregates for $r > 10^{-3}$ is beyond the scope of the present work; it will be reported separately elsewhere.

The Me-BDT molecule has two thiol groups, one at each end; whereas the insulating PT molecule has only one such thiol group, at one of its end. The thiol group in Molecular Electronics has been defined as a 'molecular alligators' due to its ability to form sulfide bond with metal electrodes. Usually bonding is an indispensable condition for the formation of ohmic contacts with the metal electrodes. The absence of one thiol group in the PT molecule leads to the formation of a spatial gap between the molecule and upper electrode. In other words Me-BDT can bond to both electrodes via sulfur-metal bonding, and thus is relatively 'transparent' to charge transport. PT molecules, however, bond only to one electrode, leading to very low conductivity (see below).

Similar molecular device structures have been fabricated previously[27,28] and several research groups have obtained useful device properties.[29,30] A counterpart to the SAM SSM technique used here is the 'nanocell' approach developed by Tour et al.[31] In the nanocell approach, disordered arrays of metallic islands are interlinked with conducting molecules. In contrast, the SAM SSM technique forms isolated molecular wires between well-defined electrodes at low r-values. Mixed monolayers of phenylethynyl thiolates diluted with alkanethiols, in which conjugated molecules exhibit higher tunneling probability through a STM tip[32-35] are most closely related to the SAM SSM molecular engineering approach used here. To the best of our knowledge there are no previous reports on molecular engineering approaches that enable a systematic study of SSM at different molecular wire surface densities. The ability to vary this density is the most significant success of our novel approach.

**B. Initial characterization of the Me-BDT/PT monolayer**

The step-by-step growth of the organic monolayers was characterized by contact angle (CA) changes, ex situ ellipsometry and UV-visible reflectance spectroscopy; these are briefly summarized in Fig. 2. After step A in Fig.1, CA changed from 17° to 45°, and the UV-Vis reflectivity spectrum of the film grown on the Au/Si substrate showed a 286 nm peak characteristic of the HOMO-LUMO transition for isolated Me-BDT molecules (Fig. 2a). Moreover the visible part of the optical reflectivity spectrum did not reveal any characteristic transition of Me-BDT molecular aggregates, which are present at large r-values. This was taken as evidence that Me-BDT molecules are indeed isolated in the PT matrix for r-values in the range $r < 10^{-4}$.

Variable angle spectroscopic ellipsometry (VASE, Woollam Co.) was used to verify the monolayer growth in the device structure.[36] The VASE measures optical spectra with 5 nm wavelength resolution in the spectral range of 300-600 nm. The structural model for fitting *ex situ* ellipsometry data uses the collected data from three different incident angles, namely 65°, 70° and 75°. The obtained and fitted ellipsometric spectra for the structures containing $Si/SiO_2$, Au, and Me-BDT/PT monolayer exhibit molecular c-axis interplanar spacing of 30.6 nm for the bottom Au film, and 10 Å for the monolayer of Me-BDT/PT SSM; this indicates single monolayer growth.

**C. Checking molecular connectivity**

Bonding with the *bottom electrode* has been well characterized in previous studies of thiol-ended SAM on various metals.[37] Aliphatic and aromatic thiolate SAM's form spontaneously on Au bottom electrode through chemisorption of the S head group to the Au surface. The monolayers interact on the surface via van der Waals forces between adjacent alkyl chains. The stability of SAM's originates from the covalent S-Au bond as well as from the attractive van der Waals forces between the adjacent molecules. As a result of the intrinsic stability of these systems, SAM's grown on metallic films are known to have low defect density, and, in addition resist degradation in air.[35] In contrast, the connectivity with the *upper electrode* is an acute problem in the field of molecular electronics.[8] To address the formation of covalent bonds between the Me-BDT molecules in the SAM SSM and the *upper Au electrode* we fabricated a SAM structure comprised of *iodopropyl-trimethoxysilane* self-assembled on a $SiO_2$/Si film. This was followed by chemisorption of either a Me-BDT monolayer (Fig. 2C), or a PT monolayer that was used as a control structure. For studying the sulfur-metal bonding of the upper electrode we used the silane matrix as a template layer for SAM (dashed arrow in Fig. 2c), thus avoiding the contribution of the bottom sulfur-metal bonding to the absorption spectra in the infrared (IR). The silane matrix is semitransparent in the mid-IR spectral range allowing absorption spectroscopy study of the upper surface vibrational modes. Upon deposition of the upper Au electrode we were able to detect the formation of Au-S bonding, because the frequency of the ir-active Au-S stretching vibration is different from that of the original C-S stretching vibration (before the metal deposition)[38] (see Fig. 2(d)). In fact the ir-active vibration frequency shifts from ~ 798 $cm^{-1}$ (Fig. 2(d) A-line) for the C-S stretching mode to ~ 614 $cm^{-1}$ (Fig. 2(d) B-line) for the Au-S mode. This red shifted mode was absent in the controlled structure that contained only PT molecules.

To validate our method we repeated the same procedure using upper cobalt electrode. Cobalt is lighter than gold leading to a smaller red shift. Indeed we found that the ir-active Co-S stretching vibrational mode shifts to 671cm$^{-1}$; the obtained shift is 127 cm$^{-1}$ compared to a shift of 184 cm$^{-1}$ for Au. This red shifted frequency is consistent with the literature data for the corresponding shift in cobalt organometallic complexes upon the formation of sulfide bonds in the case of simple flask chemistry,[38] and may be thus taken as a proof of our procedure.

**D: Counting the number of molecular wires in the device**

Counting the number of molecular wires, $N_{BDT}$ in the device is a crucial requirement for studying charge and spin transport properties of 'single' conducting molecules. The molecular conductivity may be derived from the average conductivity of many such isolated molecules in the device, divided by the number of molecules. For determining $N_{BDT}$ we used two novel detection methods and assembling strategies that were borrowed from the field of Biochemistry. These are: **(a)** surface titration of thiol groups by fluorescein-5-maleimide (F-150); **(b)** surface titration of substituted thiol groups to amino-groups by 4-nitrobenzaldehyde. These titration processes preferentially isolate the molecular wires, since there is no bonding between the active titrant molecules (or tag) and the insulating PT molecules. The basic approach for determining the number of molecular wires by surface titration is summarized in Fig. 3(a). We grew a molecular tag monolayer on top of a SAM of molecular wires (step1 in Fig. 3(a)), with a pH removable bond, having ideally1:1 ratio of tag molecules to molecular wires (step 2 in Fig. 3(a)). When changing the pH of the resulting mixture we de-assemble the tag molecules into the solution, and later determine their concentration by absorption spectroscopy (step 3 in Fig. 3(a)).

*Method (a):* For titration of the surface thiol groups, SAM SSM were prepared with various r-values, $r = 10^{-1}$, $10^{-3}$ and $10^{-5}$. The SAM surface (Fig. 3(b), step A) was then treated with a sulfur-sensitive molecular probe (Fig. 3(b)), namely F-150 Fluorescein-5 Maleimide, similar to that used in peptide research studies;[39,40] this molecule preferentially couples to the Me-BDT thiol groups at pH = 9 via the sulfhydryl (Fig. 3(b), step B). Following multiple washings in the buffer, the sample was washed multiple times in DI water, dried, and dissolved in a few drops of 37% HCl (Fig. 3(b), C). The gold layer, organic sulfide layer and attached molecular probe was then washed from the SAM with a pH = 3 buffer, and the resulting solution diluted to about 5 ml (Fig 3(b) step C). Sodium carbonate was then added to bring the pH back to ~8.5. Absorption spectroscopy of the obtained solution in the UV/Vis spectral range was performed to measure the optical density using a Cary 17 UV/Vis spectrometer. These measurements led to an estimate of the number of titrant molecules in the solution, and consequently $N_{BDT}$ in the device. For example, following a Gaussian deconvolution of the tag molecular absorption peak at 492 nm (Figure 3(c) spectrum #1) from the background (Figure 3(c) spectrum #2), we were able to estimate the OD of F150 peak in solution for r = 0.1 (Figure 3(c) spectrum #3). From this and the published molar extinction coefficient of the coupled dye molecule ($\varepsilon = 8.5 \times 10^4$ $(M \times cm)^{-1}$), we determined the molecular-wire density by counting the number of dye molecules that were adsorbed per unit area (182 $mm^2$) to be $1.5 \times 10^{10}$ molecular wires/$mm^2$. Taking into account the actual dimension of the molecular diodes (0.25 mm$^2$) and the reduction of the active device area by the shadow mask that was estimated to be ~25% of the device area, we determine $9.4 \times 10^9$ molecular wires per device for r = $10^{-1}$. Assuming a linear dilution of molecular wires in the insulating matrix we could estimate the numbers of molecular wires at smaller r-values.

The sensitive issue in the surface titration method is the assumption of 1:1 ratio of tag molecules to molecular wires. Since we cannot verify that tag molecules are bonded to all molecular wires, then the obtained value for the molecular wire density in the device is the minimum value. We note, however that if a particular molecular wire were not able to bond with the tag (due to oxidation of HS group to HSO$^-$ or for any other reason), then the same molecule would not bond either with the upper electrode; and thus would not contribute to the total device conductivity. In other words, despite some uncertainty in obtaining absolute densities associated with tag-to-wire SAM ratio, the suggested molecular titration method gives a realistic number for the molecular density in the SSM devices. In the future we plan to employ alternative methods, such as AFM imaging, electrochemistry[41] and tip enhanced fluorescent probe spectroscopy[42] for checking the reliability of our suggested surface titration approach. One such verification method is described below.

*Method (b):* Whereas titration of thiols on SAM surfaces is a new technique; in the literature there is an established method to measure the number of amino groups on SAM.[36,43] To perform this reference measurement we self-assembled 4-aminophenylthiol molecules, which are structurally similar to the Me-BDT molecules but have amino groups on the SAM structure instead of thiol groups. Following the published protocol of surface titration we estimated the density of aminobenzothiol molecules on the surface using a similar SAM SSM with r-value of $r = 10^{-1}$; we determine a density of $4.4 \times 10^8$ molecular wires per device. The two titration methods were thus in agreement with each other. We therefore conclude that the titration of surface thiols provides a useful tool for determining the density of isolated molecular wires in SAM SSM devices.

## III.     CHARGE TRANSPORT IN SAM MOLECULAR DEVICES

Following the fabrication of SAM SSM diodes on gold electrodes we have measured the I-V characteristics of the diodes at different r-values[44] and temperatures. At small r-values in the range $r < 10^{-4}$ we expect the conducting Me-BDT molecules to be isolated in the otherwise insulating PT matrix. This could be directly verified from optical reflectivity measurements that show a peak of the isolated Me-BDT molecule at about 4.2 eV (Fig. 2(a), solid line #1) and absence of any peaks in the visible spectral range that are associated with the formation of Me-BDT molecular aggregates (Fig. 2(a), dashed line #2).

Devices with r = 0

For reference, we first discuss the conductivity measurements of devices having r = 0; these are composed of insulated PT molecules with no wires. We note that all temperature dependent conductivity measurements were performed under a dynamic vacuum. The I-V curves of such a device measured at different temperatures are shown in Fig. 4(a); the detailed I-V curves in a smaller voltage interval are shown in Fig. 4(b). The I-V curves are nonlinear showing that the PT SAM device does not contain substantial amount of pinholes; otherwise it would show a linear, ohmic behavior. In addition, the I-V response curves show a dramatic temperature dependence indicating that charge injection (or extraction) via thermionic emission is dominant in these devices. In Fig. 4(c) the temperature dependent transport data are presented in terms of differential conductance spectra (DCS = dI/dV vs. V), the inset is the DCS at 15K, where the contribution of thermionic emission should be negligibly small. There is a dramatic increase in conductance of ~four orders of magnitude when the temperature changes from 100K to 300K. However below about 100K the conductance does not change as much. This is well revealed in the Arrhenius plot of ln(I) vs. 1000/T for various V's, as shown in Fig. 5(a). The estimated activation energy for this device at V = 0 is ~0.7 eV; however at higher bias voltages the activation energy decreases substantially.

Such large activation energy may be due to thermionic emission over a potential barrier that is caused by the existing gap in connectivity between the PT molecules and the upper Au electrode. In this case the DCS at T > 100K may contain anomalies at low biasing voltage that reflect the barrier height, as indeed seen in Fig. 4(c) at V ≈ 0.5 volt for 200K and 300K, respectively. Otherwise the DCS at all temperatures show an exponential increase (Fig. 4(c)) starting from a certain voltage V(on). This latter plot is especially interesting since it shows an abrupt increase at about V(on) = 5 volts. If the current is due to tunneling through the PT molecule and accompanying spatial gap between the PT molecule and Au electrode, then the DCS at 15K maps the electronic density of states of the charged molecule, which is enhanced at biasing voltages that push the electrode Fermi level, $E_F$, towards that of the HOMO (or LUMO) level of the molecule. This happens at a voltage $V(on) = 2\Delta/e$, where $\Delta$ is the energy difference between $E_F$ and the molecular HOMO (LUMO) level. With this assumption in mind we obtain for the PT molecule from $V(on) = 5$ volts $\Delta \approx 2.5$ eV, which is smaller than the HOMO-LUMO gap of this molecule.[7] This shows that the metal Fermi level lies inside the molecular gap, in agreement with its insulating transport properties.

Devices with $10^{-7} < r < 10^{-4}$

A similar analysis was also conducted for SAM SSM devices with various r-values in the range $10^{-7} < r < 10^{-4}$, where the Me-BDT molecules are isolated in the PT matrix. From the I-V curve of PT devices (r = 0) at room temperature (Fig. 4(b)) we estimated the minimum r-value at which the conductivity is dominated by the isolated Me-BDT molecules. The conductivity, $G = I/V$ is traditionally measured at small $V \sim 0.1$ volts. From Fig. 4(b) we get $G_0 = 2 \times 10^{-8}$ $\Omega^{-1}$. Any SAM SSM device having $G > 10G_0$, may then be regarded as dominated by transport through the isolated molecular wires in the device. The I-V curves of SAM SSM devices with $r = 10^{-7}$, $10^{-6}$ and $10^{-5}$ are shown respectively in Figs. 6 and 7. G at 0.1 volt for the SAM device with the smallest r-value, namely $r = 10^{-7}$ is $2 \times 10^{-6}$ $\Omega^{-1}$; this is about two orders higher than $G_0$ and it thus dominated by the molecular wires. In addition, G increases linearly with r for the other measured SSM devices, as shown in Fig. 8. We thus conclude that the lower limit r-value for which the SAM SSM devices are still dominated by transport through isolated Me-BDT molecules is $r \approx 10^{-8}$.

The conductance of the fabricated SSM device with $r = 10^{-6}$ is analyzed in more detail in Fig. 6. Fig. 6(a) shows that the nonlinear I-V characteristic is only weakly temperature dependent; in contrast to the PT device (Fig. 4). This can be also concluded from the Arrhenius plots in Fig. 5(b). The activation energy that may be extracted at intermediate biasing voltage is ~ 50 meV, which is about an order of magnitude smaller than that of the PT device (Fig. 5(a)). Since the Me-BDT molecule is bonded to the Au atoms of the two opposite electrodes, then this small activation energy cannot be due to thermionic emission over a barrier caused by a vacuum gap, as is the case for the 'PT only' device discussed above. The obtained weak temperature dependence may reflect the temperature dependence of $\Delta_{BDT}$ between $E_F(Au)$ and Me-BDT HOMO level; it is conceivable that this energy depends on the temperature, similar to many inorganic semiconductors. The weak temperature dependence may also reflect the effective molecular length, which plays an important role if the transport occurs via tunneling. In this case tunneling may be influenced by twists and/or rotation around the principal axis of the molecule, which are formed at high temperatures and thus contribute to the dependence on temperature.

A better understanding of the weak temperature dependence is provided in Fig. 6(d), where the DCS are plotted at four different temperatures. From the 15K data that should not contain any thermionic contribution, it is apparent that there is an abrupt onset voltage, V(on) for the increase in conductance at ~3 eV. As for the PT device discussed above, we can estimate $\Delta_{BDT}$ from V(on) using the relation V(on) = $2\Delta_{BDT}/e$; we get $\Delta_{BDT}$ ~1.5 eV. Fig. 6(c) also shows that V(on) decreases with the temperature indicating that indeed $\Delta_{BDT}$ depends weakly on the temperature, as assumed above.

The I-V characteristics of SSM devices with r = $10^{-7}$ and $10^{-5}$ are shown in Fig. 7 (a) and 7(c), respectively. Again I-V is highly nonlinear showing an abrupt increase at V(on) ~3 eV, similar to the device with r = $10^{-6}$ discussed above. This shows that the transport mechanism for these two devices is basically the same. Moreover the symmetry between positive and negative biasing voltages is maintained almost perfectly in all SSM devices; whereas it is less symmetric for the PT device (Fig. 4). This is in agreement with the symmetry of the fabricated SSM devices and Me-BDT molecule. In contrast, the PT molecule is less symmetric; also the gap between this molecule and the upper Au electrode may also contribute to the lack of symmetry in V for this device. The DCS for the two SSM devices are shown in Figs. 7(b) and 7(d), respectively. Once again the gap in conductance is maintained up to ~3 eV, where there is an abrupt increase in the conductance. This may show that $E_F$(Au) reaches the HOMO level at this biasing voltage. The SSM device with r = $10^{-7}$ shows a smaller gap, which may be due to the PT contribution to the conductance mechanism of this device.

We also evaluated the room temperature conductance of the SSM SAM devices at V = 0.1 volt, as depicted in Fig. 8. The linear dependence of G with r shows that the transport processes in SSM devices in this r-value range are shared by all devices up to r = $10^{-4}$, where the conductivity scales with the density of the molecular wires. We therefore conclude that charge transport in these devices is dominated by the conductance through isolated Me-BDT molecules. At higher r-values, we have measured a deviation from linearity with r, indicating the formation of Me-BDT aggregates. The formation of aggregates was independently verified by optical spectroscopies (see Fig. 3).

**IV. Single molecule resistance**

Experimental determination

The additive law of molecular devices should occur for molecular wires in parallel configuration.[4, 45] As the conductivity of the SSM diodes scales with the number of Me-BDT molecules in the device, we can extract the resistance, $R_M$ of a single molecular wire from Fig. 8. If the wires are isolated in the device then the device conductance is simply given by $\sigma = N\sigma_M$, where $\sigma_M$ is the conductance of single molecules, assuming all molecules have equivalent conductance. We may then write:

$$R = R_M/N, \qquad (1)$$

where R is the device resistance and N is the number of molecular wires in the device. From Fig. 8 and using Eq. (1) we obtained the average $R_M$ value to be $6 (\pm 3) \times 10^9$ Ω. This value is in excellent agreement with that obtained using STM measurements $R_M = 4.5 \times 10^9$ Ω,[7,46] which validates our assumptions and methods. However in contrast to STM measurements, our SSM SAM method used here can in principle be used for device application, and also enables to perform electrical measurements at low temperatures with relative ease.

Model calculation

To rationalize the measured molecular resistance $R_M$ we employed the Landauer formula for linear electrical conductance G to calculate the electrode-molecule-electrode junction resistance. In this model G is given by the relation:

$$G = 2(e^2 \times T)/h \qquad (2)$$

where h is the Planck constant, and T is the electron transmission efficiency from one contact to the other, which is a function of the applied voltage, V. T can be divided into the following three components:

$$T = T_L \times T_R \times T_M \qquad (3)$$

where $T_L$ and $T_R$ give the charge transport efficiency across the left and right contacts, and $T_M$ is the electron transmission through the molecule itself. We may approximate $T_M$ by the coherent, non-resonant tunneling through a rectangular barrier. In this case $T_M$ is given by

$$T_M = \exp(-\beta L) \qquad (4)$$

where L is the potential barrier width, i.e. the effective molecule length, and β is the tunneling decay parameter given by

$$\beta = \frac{\sqrt{2 \times m^* \alpha(\Phi - eV/2)}}{\hbar} \quad (5)$$

where $\hbar$ is h/2π, Φ is the barrier height for tunneling through the HOMO level, which is equivalent to the energy difference $\Delta_{BDT}$ between $E_F$(Au) and Me-BDT HOMO (LUMO) level, m* is the effective electron mass given in terms of $m_0$ (the free electron mass), V is the biasing voltage applied across the molecule, and α (≤1) is a parameter that describes the asymmetry in the potential profile across the electrode-molecule-electrode junction.[7]

We may estimate the electron transmission $T_M$ through the molecule using Eq. (4). The left and right electron transmission, however are more difficult to calculate. They may be negligibly small in our case since there is no charge injection barrier into the molecular channel, as indicated by the weak temperature dependent transport in our devices. The value m*/$m_0$ in Eq. (5) ranges from 0.16 for conjugated molecules, to 1.0 for saturated molecules.[7] Me-BDT has a single aromatic ring and two saturated methyl spacers, so that the value m*/$m_0$ should be intermediate between completely conjugated and completely saturated molecules. Thus we take m* = 0.58 for the Me-BDT molecule. The asymmetry parameters α should be close to 1 since the Me-BDT molecule is symmetric. Furthermore in Eq. (5) we take V = 0.1 volts and L = 10 Å (the estimated Me-BDT effective molecule length), and the tunneling barrier height Φ = $\Delta_{BDT}$ = 1.5 eV (Fig. 6). Using Eqs. (2) – (5) with the Me-BDT parameters as determined above, and neglecting the transmissions at the left and right interfaces, we obtain $R_M \approx 10^9$ Ω for the molecular resistance. This is about six times smaller than the measured $R_M$ value, but in the young field of Molecular Electronics is considered to be an excellent agreement. The very good agreement between the experimental and calculated $R_M$ values points out that thermionic emission is negligible for the Au/Me-BDT junction at room temperature, which is consistent with the weak temperature dependent transport that we have measured.

**V. Conclusions**

We explored a new molecular engineering approach for fabricating molecular devices based on isolated conducting molecules embedded in a non-conducting molecule SAM matrix. The devices employed a solid-state solution of SAM, incorporating both conducting Me-BDT and insulating PT molecules sandwiched between two Au opposite electrodes. In this configuration the Me-BDT molecules bond to both electrodes, whereas the PT molecules bond only to the bottom electrode, thereby dramatically decreasing their electrical conductivity. Following methods used in Surface Science we employed new tools to confirm connectivity of the Me-BDT with the upper Au electrode, and count the number of isolated molecular wires in the devices. We expect these methods to be applicable to a wide range of molecular engineering problems.

The electrical transport characteristics of SSM SAM diodes fabricated with different r-values of Me-BDT/PT molecule densities were studied at different temperatures. We found that a potential barrier caused by the connectivity gap between the PT molecules and the upper Au electrode dominates the transport properties of the pure PT SAM diode (r = 0). Conversely the transport properties of SSM SAM diodes having r-values in the range $10^{-8} < r < 10^{-4}$ were dominated by the conductance of isolated Me-BDT molecules in the device. The lower limit in this r-value range is determined by the finite conductance of the PT SAM matrix, whereas the upper limit is governed by the formation of Me-BDT molecular aggregates. We found that the temperature dependence of SSM SAM devices is much weaker than that of the PT SAM device, indicating the importance of molecule bonding to both electrodes. From the DCS of the various devices, we found that the energy difference, $\Delta$ between the gold electrode Fermi-level and the Me-BDT HOMO (or LUMO) level is ~1.5 eV, compared to ~2.5 eV that we found for the device based on the PT molecules. The smaller $\Delta$ value may contribute to the superior conductance of the Me-BDT molecules. We explained the weak temperature dependence of the SSM SAM devices as reflecting the weak temperature dependence of $\Delta$.

We found that the conductance of the fabricated SSM SAM devices scales linearly with r, showing that the isolated Me-BDT molecules simply add together in determining the overall device conductance. Based on this superposition, and the obtained number of the Me-BDT wire molecules in the device we determine the single molecule resistance of Me-BDT to be $R_M = 6 \times 10^8\ \Omega$. This value is in good agreement with other measurements using single molecule contact by STM spectroscopy. A simple model for calculating $R_M$, where the transport is governed by electron tunneling through the Me-BDT molecule using the WKB approximation, is in good agreement with the experimental data and thus validates the protocol followed in the present studies.

**Acknowledgements**

We acknowledge L. Wojcik help with the chemical preparation; M. Delong, X. M. Jiang and C. Yang for help with the optical measurements; J. Shi for the use of his instrumentation for measuring I-V characteristics; and B. Shapiro and M. Raikh for useful discussions. We also benefited from fruitful discussions with J. M. Gerton, L. Wojcik and A. Yakimov. This work was supported in part by the DOE Grant No. ER 46109 and NSF Grant No. 02-02790 at the University of Utah.

**Figure Captions**

**Fig. 1:** (Color on line) The fabrication process (schematic) of SAM SSM diodes at small ratio r of molecular wire (Me-BDT in red) to molecular insulator (PT in green). A: evaporation of an Au base electrode; B: SAM growth of the appropriate molecule mixture on the bottom Au film; C: evaporation of the upper Au electrode; D: I-V measurement set-up, where the contacts are made via silver paint.

**Fig. 2:** (Color on lone) Basic optical measurements of the SAM devices. (a) The optical reflectivity spectrum of a SAM film with $r = 10^{-4}$ that shows a prominent feature at the HOMO-LUMO transition of the isolated Me-BDT molecule (blue solid line #1). There is no other optical feature in the visible spectral range, indicating the lack of aggregate formation. Aggregate peak that occurs at high r-values (here $r = 10^{-2}$) is shown as a reference (red dashed line #2). (b) Spectra of the two optical constants, $\psi$ (blue) and $\Delta$ (red) used in ellipsometry that are measured at three different angles, from which a film thickness of ~1 nm was derived (green line is model fitting). (c) Schematic representation of the method used to verify Me-BDT connectivity to the upper Au electrode. The dashed red line corresponds to AuS interface bond. (d) FTIR absorption spectra of the Me-BDT molecule bonded to the Au electrode that shows an ir-active AuS-C stretching vibration (blue, B), compared with a reference film that shows the ir-active S-C stretching vibration (red, A).

**Fig. 3:** (Color on line) Schematic representation of the titration method used to count the molecular wires in the SSM SAM devices (a) and (b) and the absorption spectrum of the product titrant molecular tag in solution (c). (a) The steps 1, 2 and 3 (self-explanatory) that lead to the molecular tags in solution. (b) A more detailed explanation of the titration process; steps A, B, and C are assigned. Symbol R on the last scheme corresponds to molecular wire that could be attached to molecular tag. Since experimentally $\lambda_{max}$ of tag molecule does not effected by R, then $\varepsilon$ of tag likely remains the same, disregarding actual R nature (c) The absorption spectrum of the tag molecules in solution following the titration process of a SAM grown with r = 0.1. The self-explanatory steps 1, 2, and 3 are assigned.

**Fig. 4:** (Color on line) Electrical transport studies of a SAM device made of PT (r = 0) at various temperatures. (a) and (b) show the measured I-V characteristics; (c) show the differential conductivity spectra obtained from (a). Insert in (c) is dI/dV vs. V for 15 K.

**Fig. 5:** (Color on line) Arrhenius plots of the current at different biasing voltages for the SAM device of PT (data taken from Fig. 4) (a); and SSM SAM device with r = $10^{-6}$ (data taken from Fig. 6) (b). Red lines are linear fits for ln(I) rise at high temperatures.

**Fig. 6:** (Color on line) Same as in Fig. 4, but for a SSM SAM device with r = $10^{-6}$.

**Fig. 7:** (Color on line) Same as in Fig. 4 but for SSM SAM devices at room temperature with r = $10^{-5}$ (a) and (b); and r = $10^{-7}$ (c) and (d).

**Fig. 8**: (Color on line) Room temperature current of SSM SAM devices fabricated with different r-values vs. r. A linear line through the data points is also shown indicating the dominant role of the Me-BDT conductivity superposition.

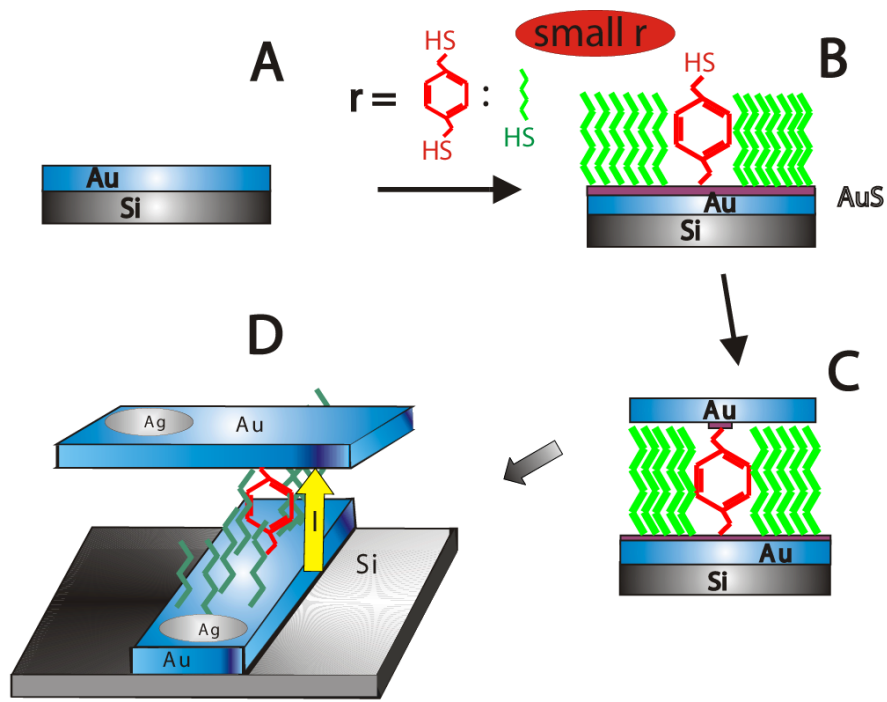

Fig. 1.

(a)
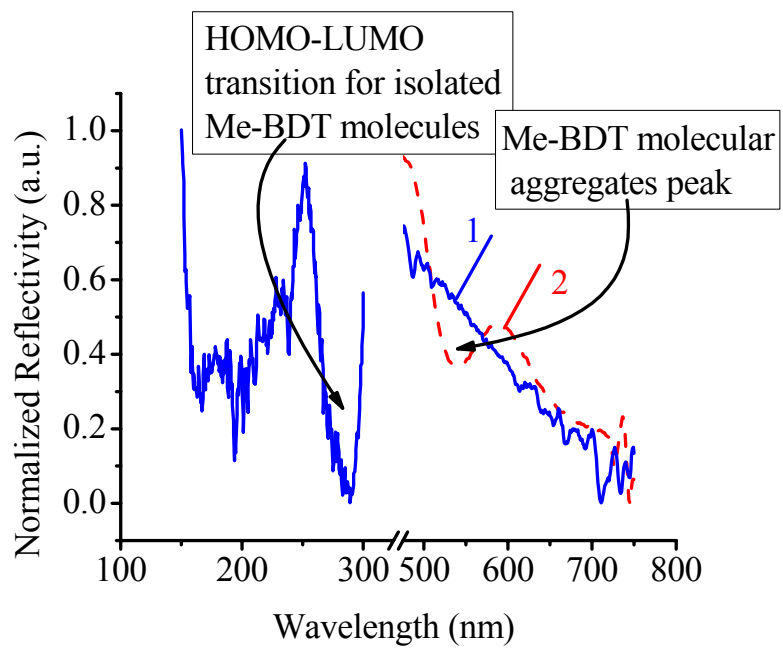

Fig. 2.

(b)

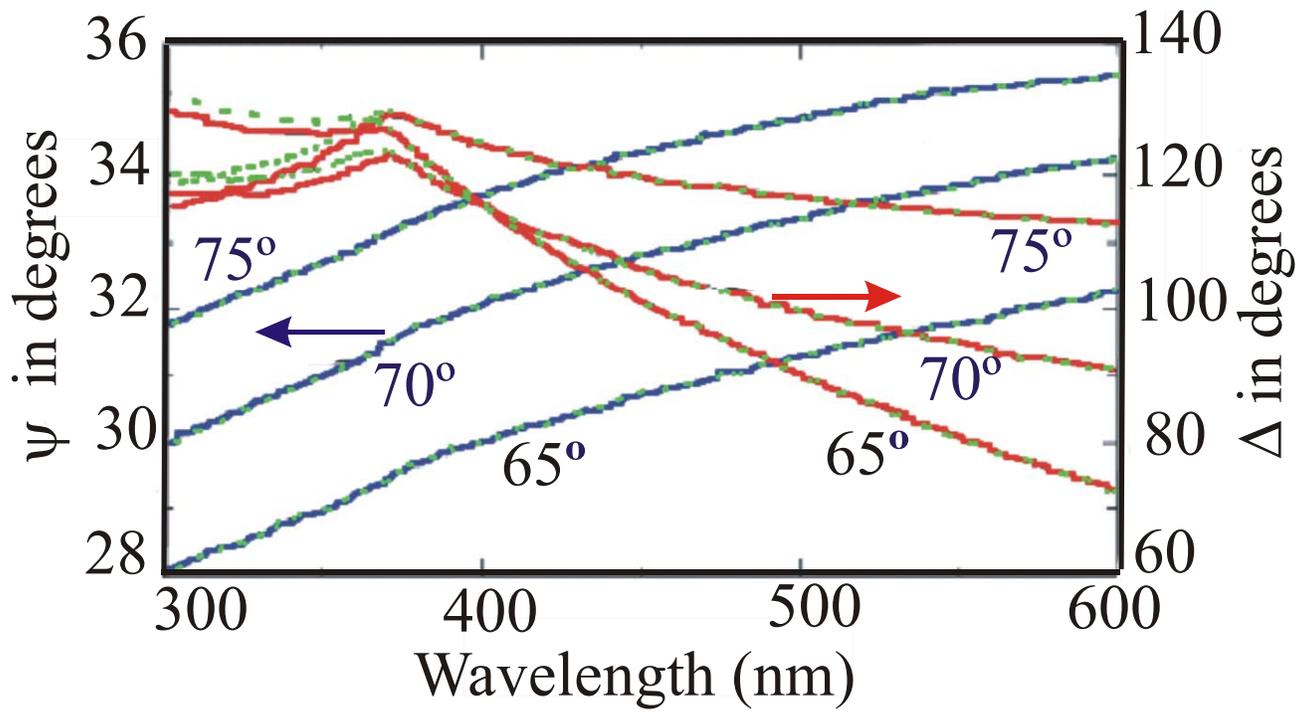

(c)

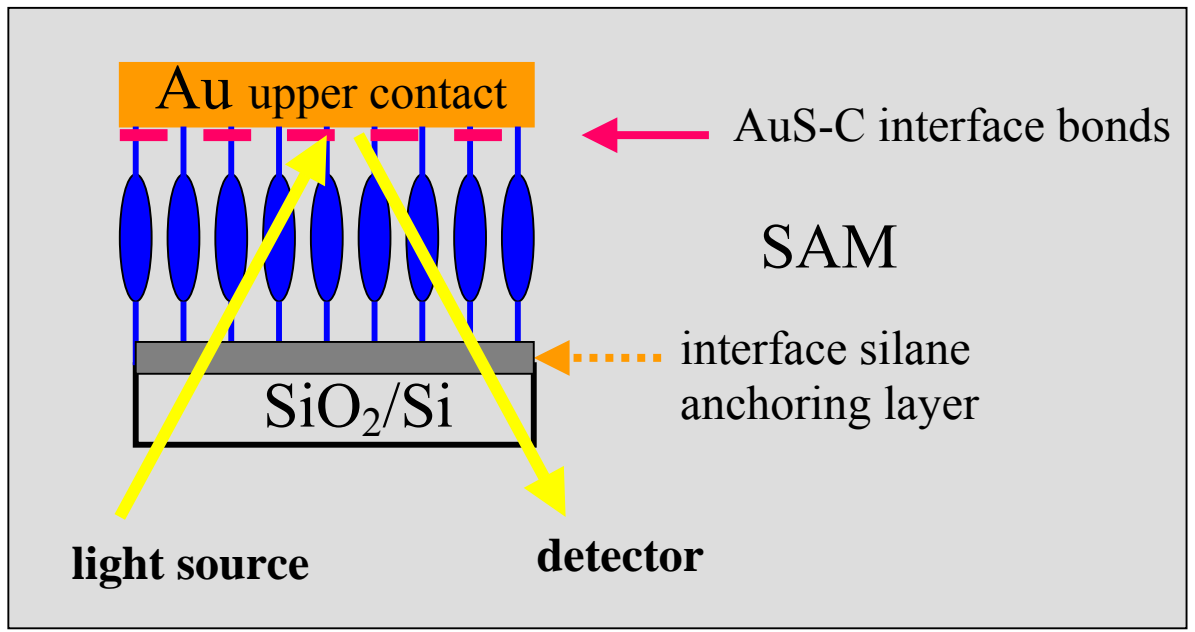

Fig. 2.

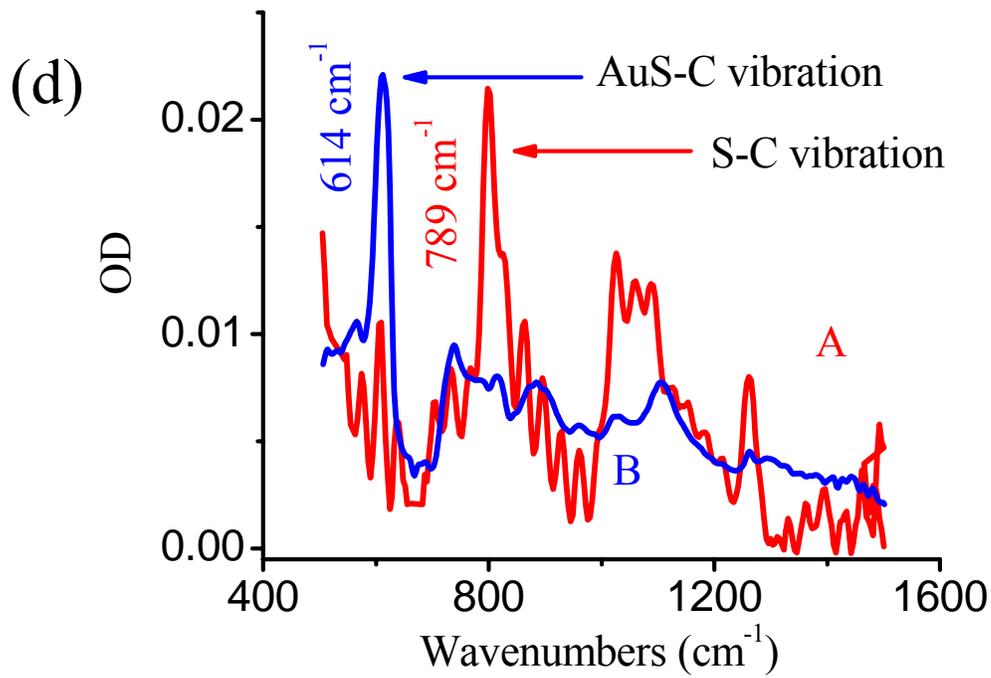

Fig. 2.

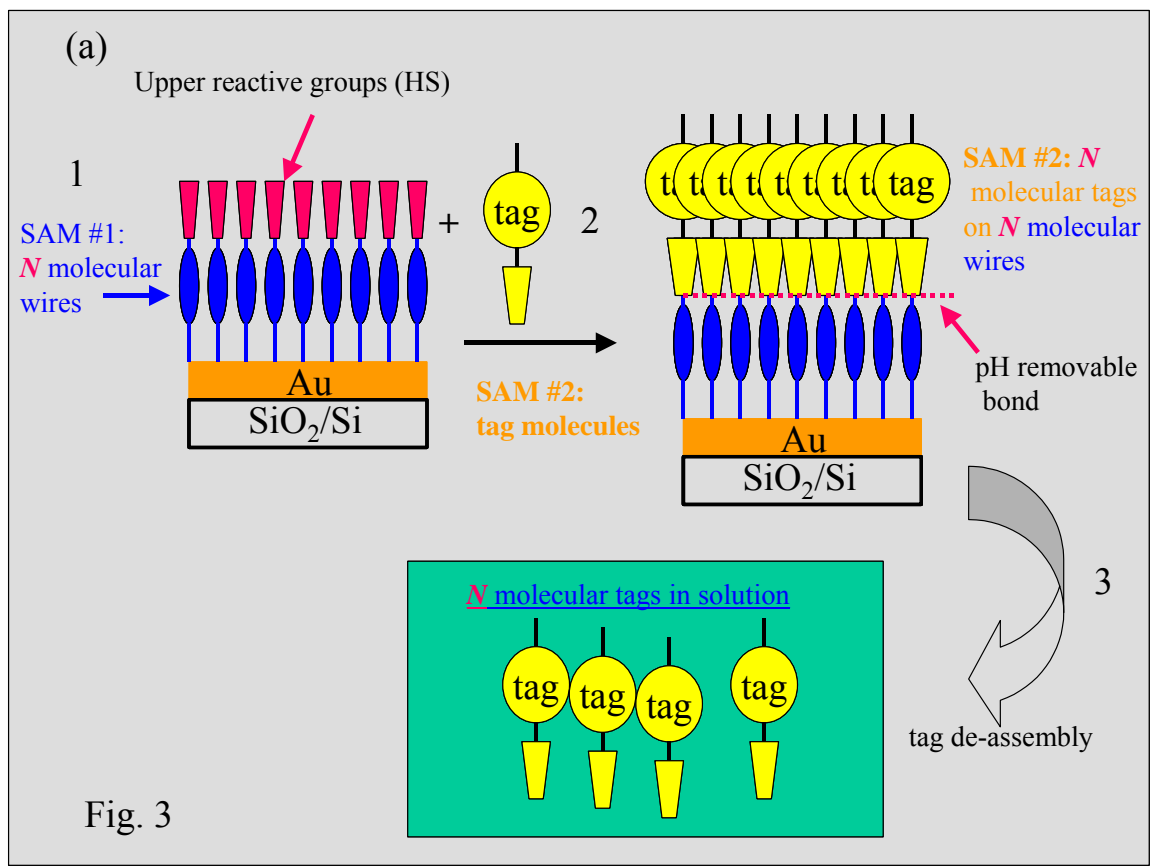

Fig. 3

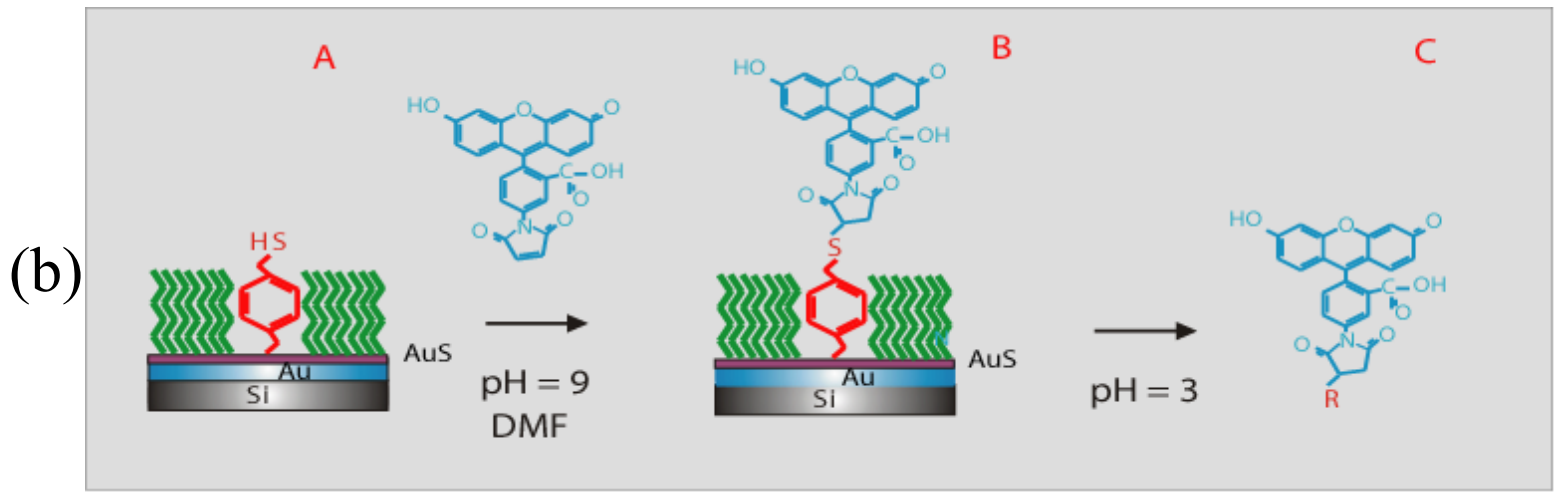

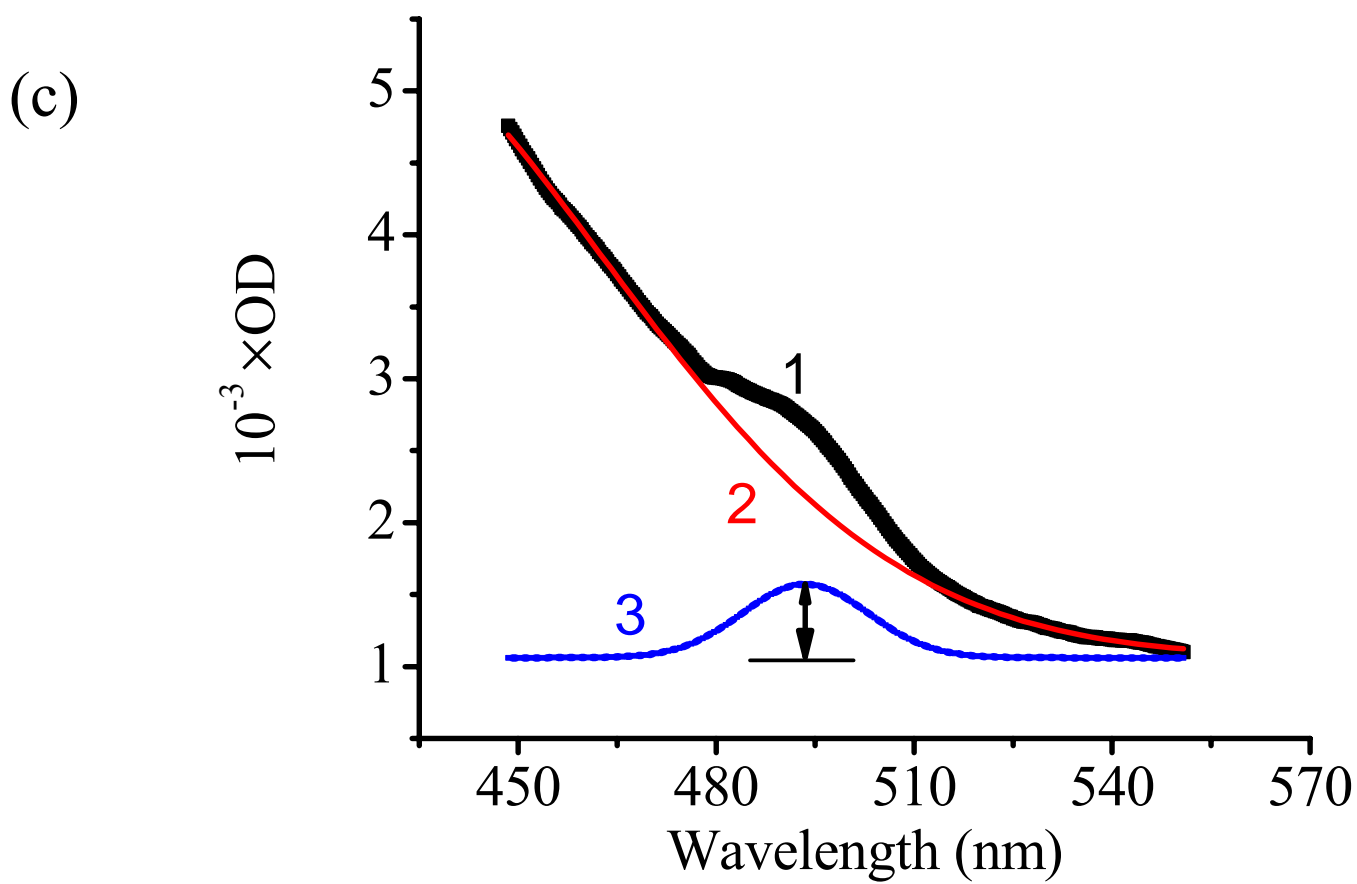

Fig. 3

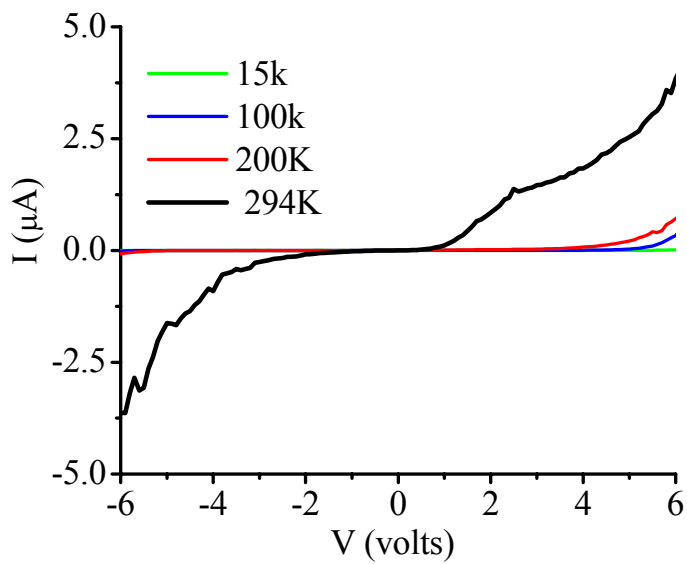
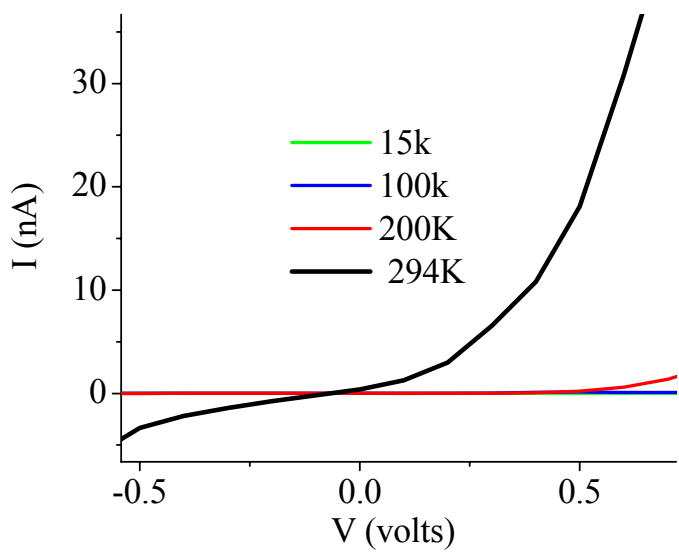
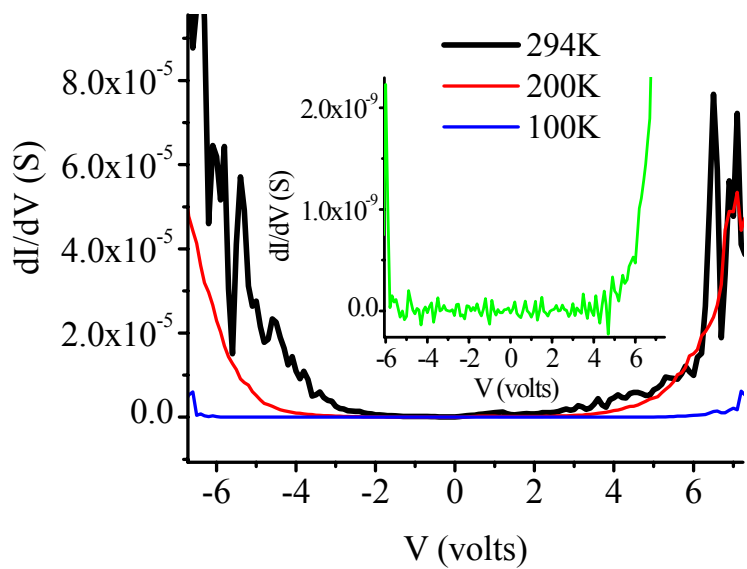

Fig. 4

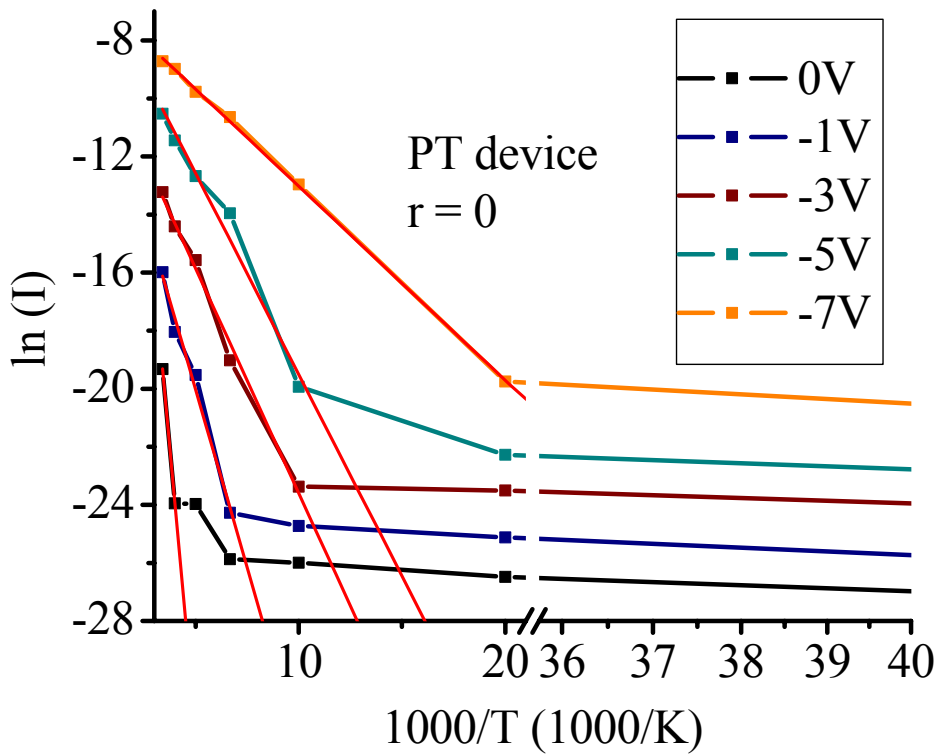

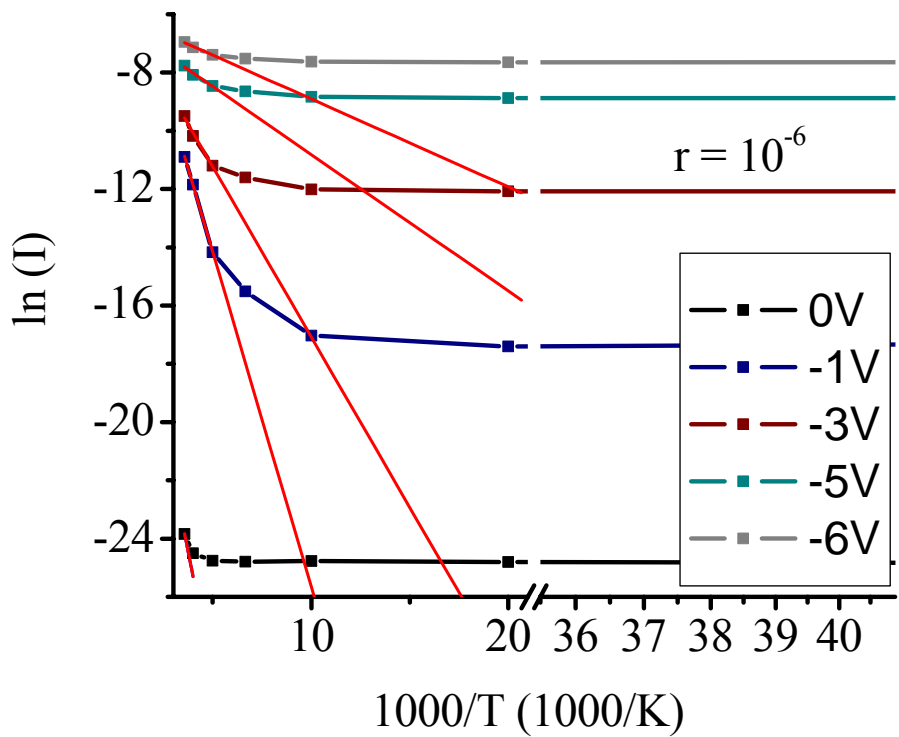

Fig. 5

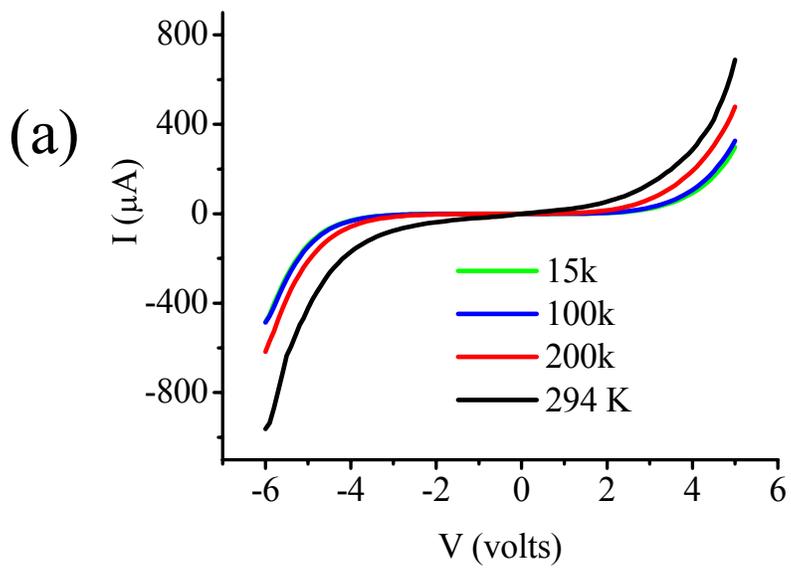
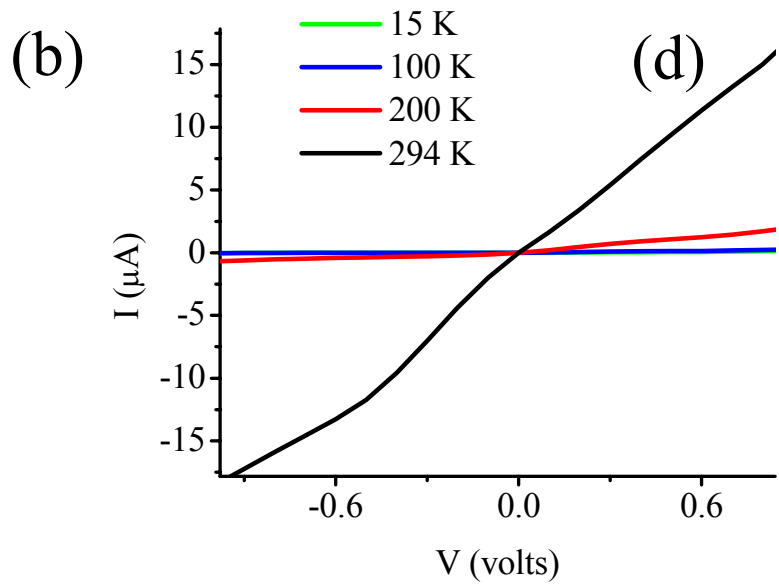
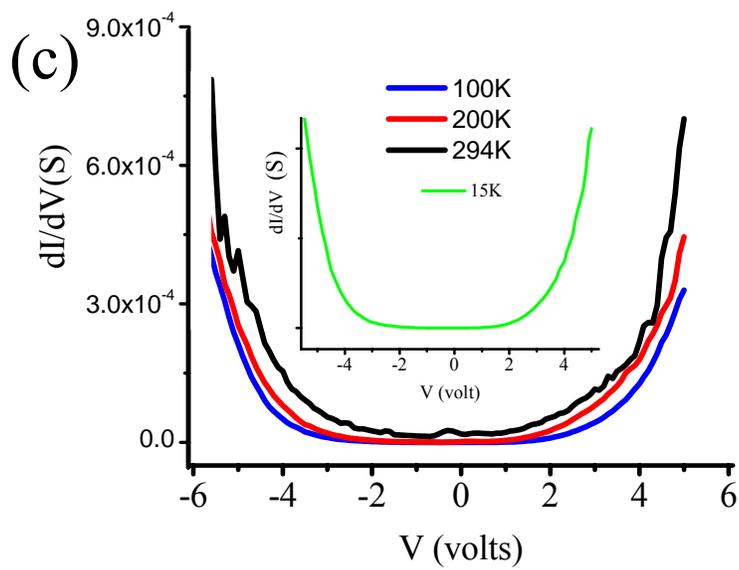

Fig. 6

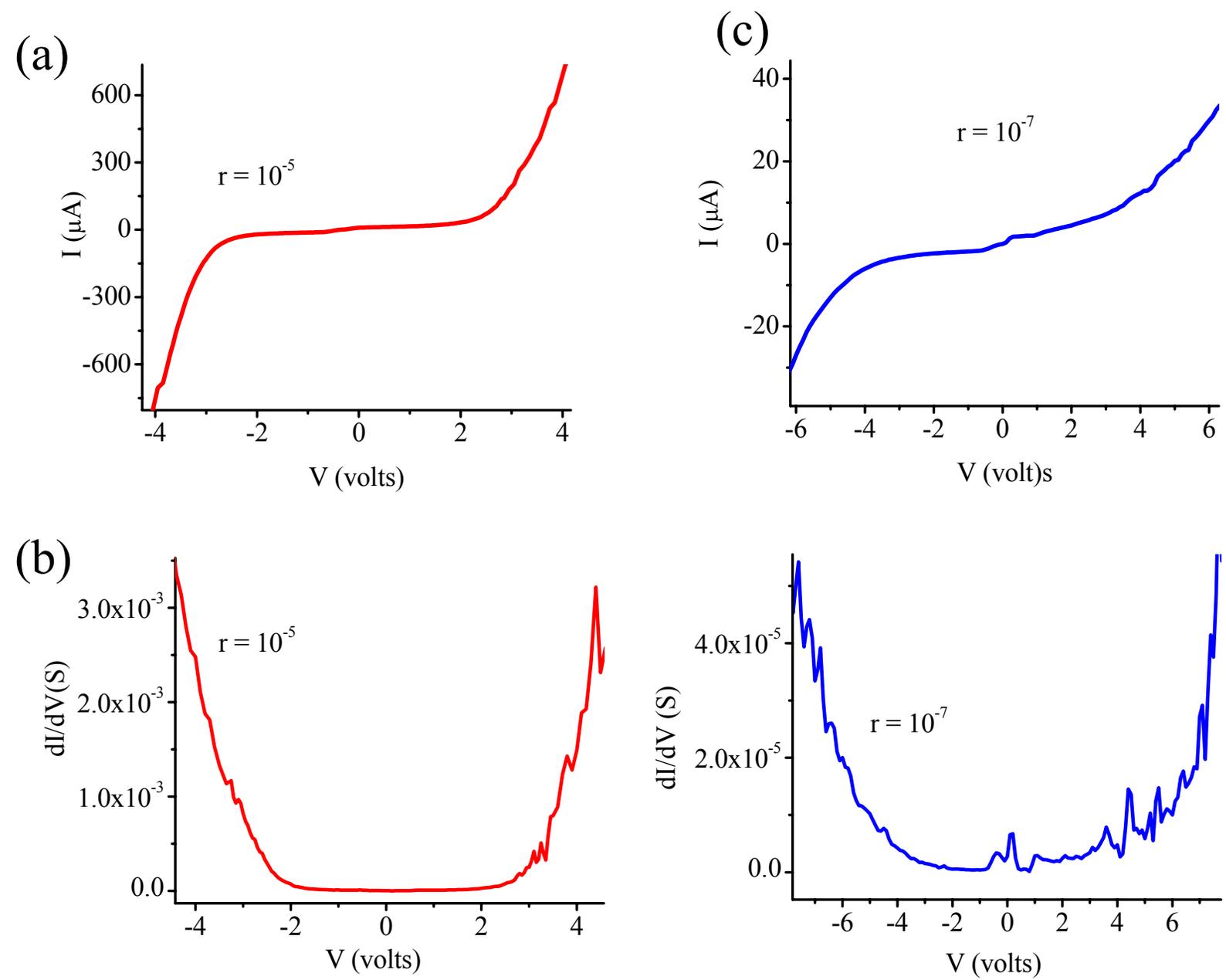

Fig. 7.

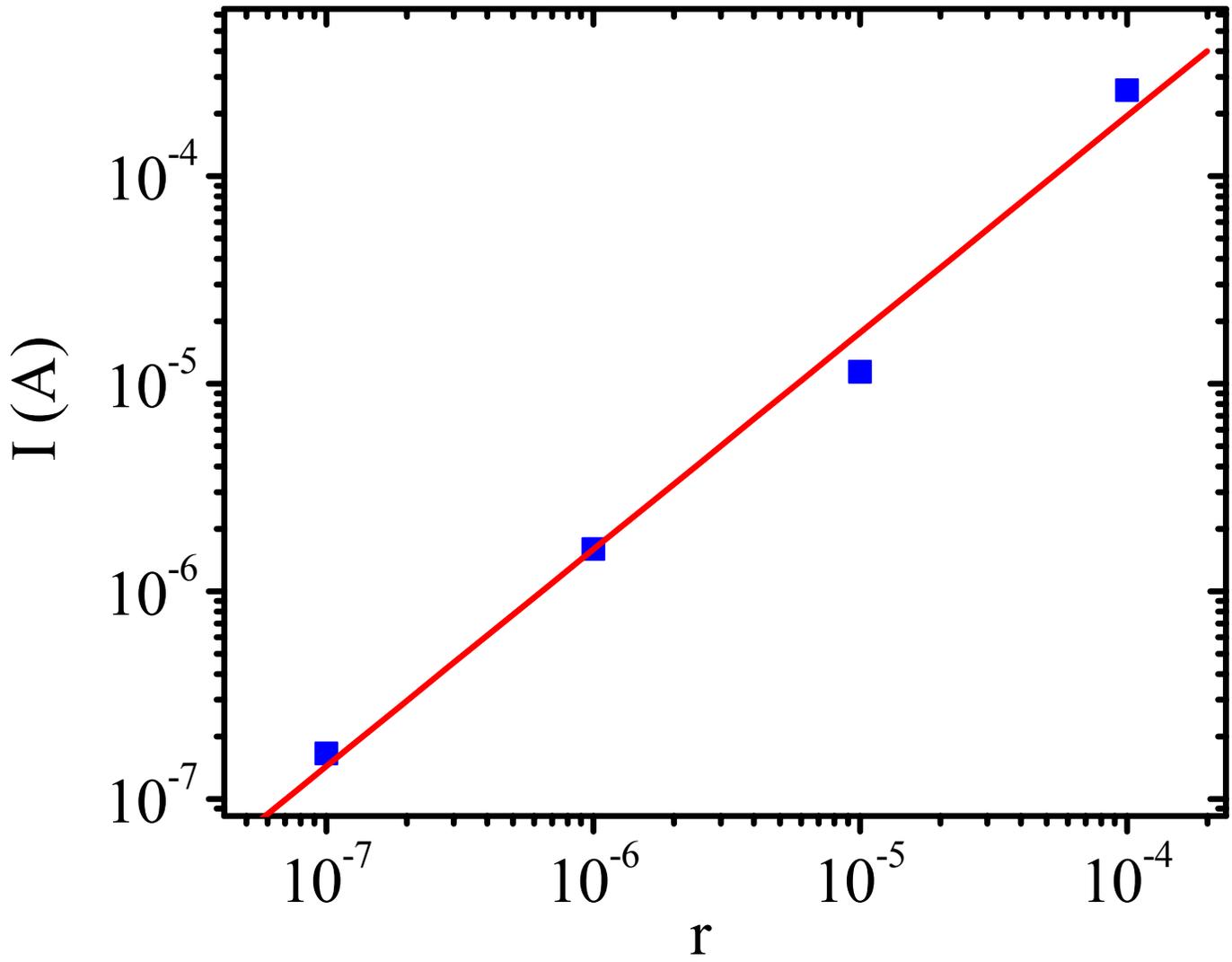

Fig. 8